\documentclass{article}
\usepackage{amssymb}
\usepackage{amsmath}
\usepackage{latexsym}
\usepackage{array}
\usepackage{graphicx}
\usepackage{txfonts}
\usepackage{theorem}
\theoremstyle{break}
\newtheorem{Theorem}{Theorem}

\newtheorem{Proposition}{Proposition}
\newtheorem{Corollary}{Corollary}

\def\qed{\hfill\hbox{$\Box$}\vspace{10pt}\break}

\def\Z{{\mathbb Z}}

\def\R{{\mathbb R}}

\begin{document}
\title{\Large{Soliton Solutions of a Generalized Discrete KdV Equation}}
\author{Masataka Kanki$^1$, Jun Mada$^2$ and Tetsuji Tokihiro$^1$\\ \\
$^1$ Graduate school of Mathematical Sciences,\\
      University of Tokyo, 3-8-1 Komaba, Tokyo 153-8914, Japan\\
$^2$College of Industrial Technology,\\
  Nihon University, 2-11-1 Shin-ei, Narashino, Chiba 275-8576, Japan}
\date{}
\maketitle

\begin{abstract}
We investigate the multi-soliton solutions to the generalized discrete KdV equation. In some cases a soliton with smaller amplitude moves faster than that with larger amplitude unlike the soliton solutions of the KdV equation.
This phenomenon is intuitively understood from its ultradiscrete limit, where the system turns to the box ball system with a carrier.\\
\tt{Keywords: soliton, integrable equation, nonlinear system, discrete KdV equation, cellular automaton}
\end{abstract}
\section{Introduction}
\label{sec1}

The discrete KdV equation
\begin{equation}
\frac{1}{x_{n+1}^{t+1}}-\frac{1}{x_n^t}+\frac{\delta}{1+\delta}\left(x_n^{t+1}-x_{n+1}^t \right)=0
\label{dKdV1}
\end{equation}
is an integrable partial difference equation\cite{HirotaTsujimoto}.
Here $n,t \in \Z$ and $\delta\in \R$ is a parameter.
The eq. \eqref{dKdV1} turns to the continuous KdV equation by taking an appropriate continuous limit and has multi-soliton solutions.
The solitons in \eqref{dKdV1} have similar properties to those in continuous KdV equation.
They do not change their amplitude after collision and a soliton with larger amplitude moves faster than that with smaller amplitude. Note that when we put
\[
\frac{1}{y_n^t}:=(1+\delta)\frac{1}{x_n^{t+1}}-\delta x_n^t
\]
we obtain equivalent coupled equations 
\begin{equation}
\left\{
\begin{array}{cl}
x_n^{t+1}&=\dfrac{(1+\delta)y_n^t}{1+\delta x_n^ty_n^t},\vspace{2mm} \\
y_{n+1}^{t}&=\dfrac{(1+\delta x_n^ty_n^t)x_n^t}{1+\delta}.
\end{array}
\right.
\label{dKdV2}
\end{equation}
In order for us to properly define the evolution of the eq. \eqref{dKdV2}, we impose a boundary condition
$x_n^t=y_n^t=1$ for all $n\le n_0$, where $n_0$ is a small negative integer. We impose this condition on other systems in this article whenever it is necessary.  
In 1990, Takahashi and Satsuma \cite{TS} proposed a cellular automaton with soliton solutions which
is now called a `box ball system' (BBS). The time evolution rule of BBS is as follows:
\begin{equation}
U_n^{t+1}=\min\left[C_B-U_n^t,\ \sum_{i=-\infty}^{n-1}\left(U_i^t-U_i^{t+1}\right)\right] \label{BBS0}
\end{equation}
where $U_n^t$ is a number of balls in $n$th box at time $t$ and $C_B$ means every box holds at
most $C_B$ balls. The parameter $C_B$ is called `capacity of the box'.
We introduce another variable $V_n^t$ to obtain coupled equations which are equivalent to the BBS \eqref{BBS0}:
\begin{equation}\label{BBS1}
\left\{
\begin{array}{cl}
U_n^{t+1}&=\min\left[C_B-U_n^t,\ V_n^t\right],\vspace{2mm}\\
V_{n+1}^t&=U_n^t+V_n^t-U_n^{t+1}.
\end{array}
\right.
\end{equation}
This system is known to be derived from the discrete KdV eq. \eqref{dKdV1} by a limiting procedure called the ultradiscretization\cite{TTMS}. 
The BBS also has multi-soliton solutions.
A peculiar feature of solitons in BBS is that the velocity of a soliton is exactly proportional to its amplitude, the number of balls constituting the soliton, just like a soliton in the continuous KdV equations.  

\begin{figure}
\centering
\includegraphics[width=8cm,bb=40 -40 500 460]{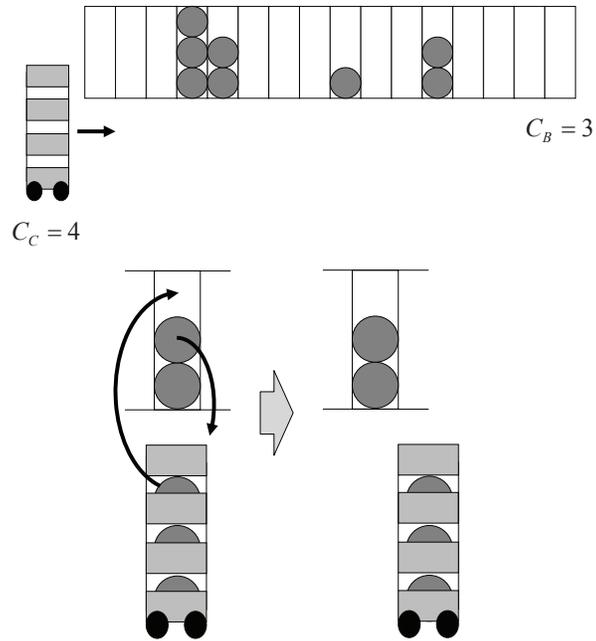}
\caption{The action of the carrier exchanging the balls with the box. From time $t$ to $t+1$, the carrier passes each box from the left to the right. Note that getting the balls from the box and putting the balls to the box take place at the same time.}
\label{figure4}
\end{figure}

\begin{figure}
\centering
\includegraphics[width=8cm,bb=-50 50 430 480]{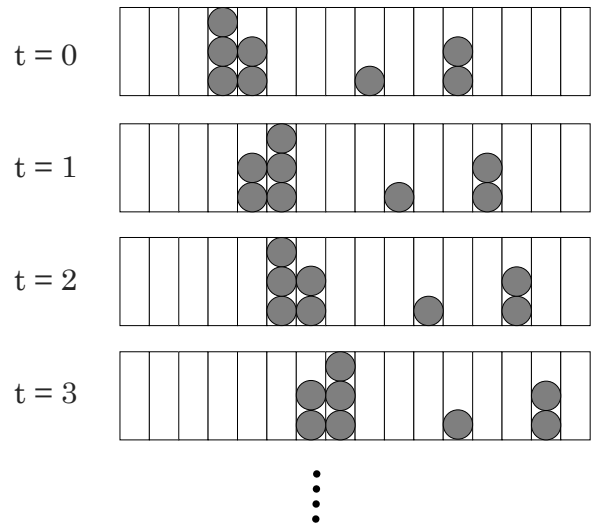}
\caption{The time evolution of the BBSC for $C_B=3$ and $C_C=4$.}
\label{figure5}
\end{figure}

The `box ball system with a carrier' (BBSC) \cite{TM} is a generalization of the BBS defined as follows:
\begin{equation}\label{BBSC}
\left\{
\begin{array}{cl}
U_n^{t+1}&=\min\left[C_B-U_n^t,\ V_n^t\right]+\max\left[0,\ U_n^t+V_n^t-C_C\right],\vspace{2mm}\\
V_{n+1}^t&=U_n^t+V_n^t-U_n^{t+1},
\end{array}
\right.
\end{equation}
where $C_C$ is a positive parameter which is called `capacity of the carrier'.
To describe the evolution rule of the BBSC we prepare a `carrier' of balls.
We assume that the carrier can carry at most $C_C$ balls.
From time $t$ to $t+1$, carrier moves from the $-\infty$ site to the $\infty$ site and passes each box from the left to the right. The carrier gets as many balls as possible from the $n$th box and, at the same time puts as many balls as possible to the vacant space of the $n+1$th box for $n=\cdots,-2,-1,0,1,2,\cdots$. The action of the carrier and the time evolution of BBSC for $C_B=3$ and $C_C=4$ is described in figs. \ref{figure4} and \ref{figure5}.

In this paper we treat a generalized system of \eqref{dKdV2} and calculate the velocity and the amplitude of the soliton solution to the equation. We present several examples of these solutions. The ultradiscrete limit of the generalized equation we treat in this article is proved to be equivalent to the BBSC.
One interesting property of the solitons in the generalized system is that a soliton with smaller amplitude moves faster than that with larger amplitude in some parameter region.
This phenomenon is intuitively understood by the corresponding solitons in the BBSC.

\section{Generalized Discrete KdV Equation and its $N$-soliton Solutions} 
\begin{figure}
\centering
\includegraphics[width=12cm,bb=50 420 480 750]{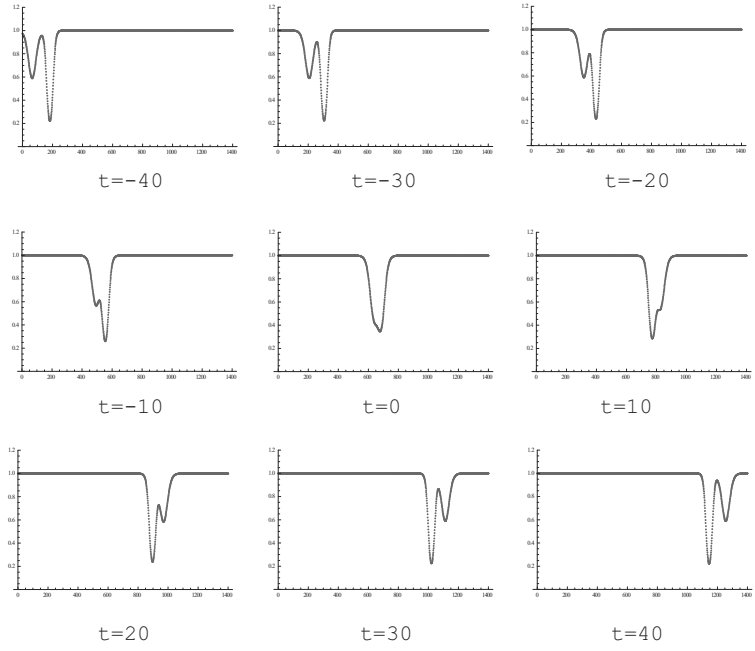}
\caption{The two-soliton solution of the generalized discrete KdV equation where $\alpha=5/6<\beta=14/15$. We observe that
the smaller soliton takes over the larger one.}
\label{figure1}
\end{figure}

\begin{figure}
\centering
\includegraphics[width=12cm,bb=100 550 470 750]{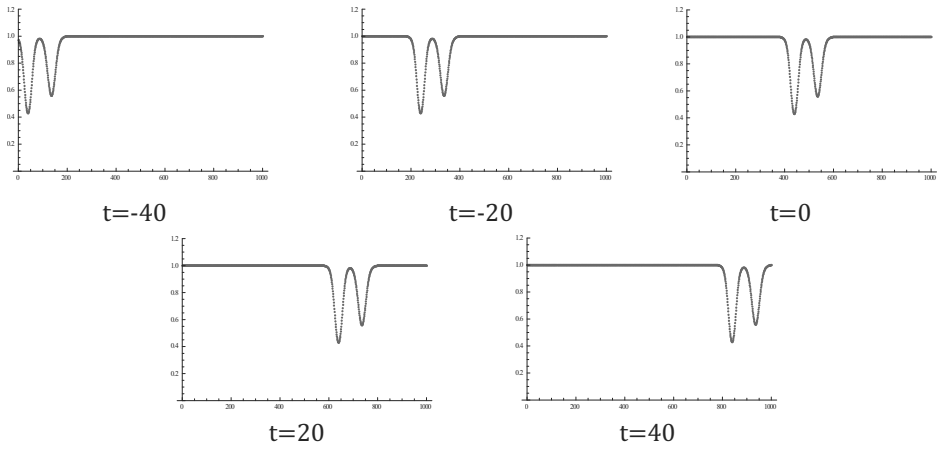}
\caption{The solution of the generalized discrete KdV equation where $\alpha=\beta=5/6$. Every point moves at speed one if $\alpha=\beta$.}
\label{figure2}
\end{figure}

Let us consider a generalized discrete KdV equation\cite{KMT}
\begin{equation}
\left\{
\begin{array}{cl}
x_n^{t+1}&=\dfrac{(1-\beta)+\beta x_n^t y_n^t}{(1-\alpha)+\alpha x_n^t y_n^t}y_n^t,\vspace{2mm}\\
y_{n+1}^t&=\dfrac{(1-\alpha)+\alpha x_n^t y_n^t}{(1-\beta)+\beta x_n^t y_n^t}x_n^t.
\end{array}
\right.
\label{YBdKdV}
\end{equation}
By a scaling transformation
\begin{equation}
x_n^t:=(1-\beta) u_n^t,\ y_n^t:=(1-\alpha) v_n^t, \label{restr}
\end{equation}
and change of variables $a=\alpha(1-\beta)$ and $b=\beta(1-\alpha)$,
the eq. \eqref{YBdKdV} becomes the 3D-consistency condition for the discrete potential KdV eq., which is an example of the Yang-Baxter map \cite{KakeiNimmoWillox, PTV},
\begin{equation}\label{YB2}
R(b,a):\  \left\{
\begin{array}{cl}
u_n^{t+1}&=\dfrac{(1+bu_n^t v_n^t)v_n^t}{1+au_n^t v_n^t},\vspace{2mm} \\
v_{n+1}^t&=\dfrac{(1+au_n^t v_n^t)u_n^t}{1+ bu_n^t v_n^t}.
\end{array}
\right.
\end{equation}
By putting $\zeta_n^t=\sqrt{a(b+1)} u_n^t$, $\xi_n^t=\sqrt{\dfrac{ab^2}{b+1}} v_n^t$ and $\delta=b^{-1}$,we obtain
\begin{equation}
\left\{
\begin{array}{cl}
\zeta_n^{t+1}&=\dfrac{(1+\delta)\xi_n^t}{1+\delta \zeta_n^t \xi_n^t}\cdot\left(1+\dfrac{\zeta_n^t \xi_n^t}{a}\right),\vspace{2mm}\\
\xi_{n+1}^t&=\dfrac{(1+\delta \zeta_n^t \xi_n^t)\zeta_n^t}{(1+\delta)}\cdot\left(1+\dfrac{\zeta_n^t \xi_n^t}{a}\right)^{-1}.
\end{array}
\right.\label{dKdVlim}
\end{equation}
By taking the limit $a\to+\infty$, the eq. \eqref{dKdVlim} becomes \eqref{dKdV2} and the eq. \eqref{YBdKdV} is indeed a generalization of the discrete KdV eq. \eqref{dKdV2}.
The eq. \eqref{YBdKdV} is known to be derived from the reduction of the discrete KP equation.
\begin{Proposition}[Date-Jimbo-Miwa\cite{DJKM}]
Let us consider 4 component discrete KP equation:
\begin{align}
&(a_1-b)\tau_{l_1t}\tau_n+(b-c)\tau_{l_1}\tau_{tn}+(c-a_1)\tau_{l_1n}\tau_t=0,
\label{eq1}\\
&(a_2-b)\tau_{l_2t}\tau_n+(b-c)\tau_{l_2}\tau_{tn}+(c-a_2)\tau_{l_2n}\tau_t=0.
\label{eq2}
\end{align}
Here $\tau=\tau(l_1,l_2,t,n)$ $((l_1,l_2,t,n) \in \Z^4)$ is the $\tau$-function, and $a_1,\ a_2,\ b,\ c$ are arbitrary parameters and we use the abbreviated form,
$\tau \equiv \tau(l_1,l_2,t,n),\  \tau_{l_1} \equiv \tau(l_1+1,l_2,t,n),\ \tau_{l_1t} \equiv \tau(l_1+1,l_2,t+1,n) \ $
and so on.
Then the $N$-soliton solution to eqs. \eqref{eq1} and \eqref{eq2} is
\begin{equation}
\tau=\det_{1\le i,j \le N}\left[ \delta_{ij}+\frac{\gamma_i}{p_i-q_j}\left(\frac{q_i-a_1}{p_i-a_1 }\right)^{l_1}
\left(\frac{q_i-a_2}{p_i-a_2 }\right)^{l_2}\left(\frac{q_i-b}{p_i-b }\right)^{t}\left(\frac{q_i-c}{p_i-c }\right)^{n} \right], \label{4kpsol}
\end{equation}
where $\{p_i,q_i\}_{i=1}^N$ are the parameters distinct from each other and $\{ \gamma_i \}_{i=1}^N$ are
arbitrary parameters.
\end{Proposition}
Using this fact we obtain the $N$-soliton solutions to \eqref{YBdKdV} by a similar reduction adopted in Kakei-Nimmo-Willox \cite{KakeiNimmoWillox}.
\begin{Proposition}[Kanki-Mada-Tokihiro\cite{KMT}]\label{fact2}
The $N$-soliton solutions of \eqref{YBdKdV} are
\begin{equation}
x_n^t:=\frac{f g_n}{g f_n},\qquad y_n^t:=\frac{g f_t}{f g_t}
\label{xy}
\end{equation}
where
\begin{align}
f&=\det_{1\le i,j \le N}\left[ \delta_{ij}+\frac{\gamma_i}{p_i+p_j+\Delta }\left(\frac{-p_i+\beta}{p_i+1-\alpha }\right)^{t} 
\left(\frac{p_i+1-\beta}{-p_i+\alpha}\right)^{n} 
\right], \label{fform}\\
g&=\det_{1\le i,j \le N}\left[ \delta_{ij}+\frac{\gamma_i}{p_i+p_j+\Delta }\frac{-\Delta -p_i}{p_i}\left(\frac{-p_i+\beta}{p_i+1-\alpha }\right)^{t} 
\left(\frac{p_i+1-\beta}{-p_i+\alpha}\right)^{n}
\right], \label{gform}
\end{align}
with $\Delta=1-\alpha-\beta$.
\end{Proposition}
\textbf{Sketch of the proof}\quad
Imposing the reduction condition
\begin{equation}
\tau_{l_1l_2}=\tau, \label{reduction}
\end{equation}
to \eqref{eq1} and \eqref{eq2} gives the constraint
\[
\left(\frac{a_1-p_i}{a_1-q_i}\right)\left(\frac{a_2-p_i}{a_2-q_i}\right)=1
\]
to the parameters $\{p_i,\, q_i\}$ in \eqref{4kpsol}.
Since $p_i \ne q_i$, the constraint becomes $p_i+q_i=a_1+a_2$.
Then we define $f:=\tau,\ g:=\tau_{l_1}$ from the solution to \eqref{eq1} and \eqref{eq2}.
Putting  $\alpha:=\dfrac{c-a_1}{c-b},\ \beta:=\dfrac{a_2-b}{c-b}$,
we find that $x_n^t,\ y_n^t$ defined by \eqref{xy} satisfy the eq. \eqref{YBdKdV}.
By redefining the parameters as $\dfrac{p_i-a_1}{c-b}  \rightarrow  p_i$,
$\dfrac{\gamma_i}{c-b} \rightarrow  \gamma_i  $ and by putting $l_1=l_2$ we have the result.
\qed
Let us consider the one-soliton solution \eqref{fform} and \eqref{gform} for $N=1$, $p_1=p$ and $\gamma_1=\gamma$.
If we take
\[
A=\frac{-p+\beta}{p+1-\alpha},\ B=\frac{p+1-\beta}{-p+\alpha},\ C=\frac{\gamma}{2p+\Delta},\ D=\frac{-\Delta-p}{p},
\]
the solution $x_n^t$ of the eq. \eqref{YBdKdV} has the  following form:
\[
x_n^t=\frac{(1+CA^t B^n)(1+CDA^tB^{n+1})}{(1+CDA^tB^n)(1+CA^tB^{n+1})}.
\]
We assume that the parameters satisfy the condition
\[
0<\alpha<1,\ 0<\beta<1,\ 0<p<\alpha+\beta-1,\ 0<\gamma \cdot \left(p-\frac{\alpha+\beta-1}{2}\right),
\]
to assure that we have $A>0,\ B>0,\ C>0$ and $D>0$, which are sufficient to obtain a one-soliton with a bounded amplitude.
The velocity $v(p)$ of $x_n^t$ is
\begin{equation}
v(p):=
\left\{
\begin{array}{cl}
-\dfrac{\log A}{\log B} & p\neq \dfrac{\alpha+\beta-1}{2},\\
1 & p=\dfrac{\alpha+\beta-1}{2}.
\end{array}
\right.
\end{equation}
We then calculate the  fluctuation of  $x_n^t$ with respect to $X:=B^n$. With some parallel displacement with respect to $t$ we can assume that $CA^t=1$ without changing the amplitude of the solution. We obtain
\[
\frac{dx_n^t}{dX}=(B-1)(D-1)\frac{1-BDX^2}{(1+BX)^2(1+DX)^2}.
\]
Note that
\begin{equation*}
(B-1)(D-1)
\left\{
\begin{array}{cl}
<0 & p \neq -\dfrac{\Delta}{2},\vspace{2mm}\\
=0 & p=-\dfrac{\Delta}{2}.
\end{array}
\right.
\end{equation*}
Therefore $x_n^t$ has a local minimum at $X=\dfrac{1}{\sqrt{BD}}$ if $p \neq -\dfrac{\Delta}{2}$, and $x_n^t=1$ for all $X$ if $p=-\dfrac{\Delta}{2}$.
Summing up the results obtained above, the amplitude $W(p)$ of one-soliton $x_n^t$ is as follows:
\begin{equation}
W(p)=\left|x_{n_0}^t-1\right|=\left|\frac{\left(1+\frac{1}{\sqrt{BD}}\right)\left(1+\sqrt{BD}\right)}{\left(1+\sqrt{\frac{D}{B}}\right)\left(1+\sqrt{\frac{B}{D}}\right)}\right|, \label{width}
\end{equation}
where $n_0$ satisfies $B^{n_0}=\dfrac{1}{\sqrt{BD}}$.
\begin{Theorem}
For the velocity $v(p)$ and the amplitude $W(p)$ of one-soliton solution of the generalized KdV eq. \eqref{YBdKdV} we have the following properties:
\begin{itemize}
\item $W(p)$ is monotone decreasing on $[0,(\alpha+\beta-1)/2]$ and
monotone increasing on $[(\alpha+\beta-1)/2,\alpha+\beta-1]$.
\item If $0<\alpha<\beta<1$ then $v(p)$ is monotone increasing on $[0,(\alpha+\beta-1)/2]$ and
monotone decreasing on $[(\alpha+\beta-1)/2,\alpha+\beta-1]$.
\item If $0<\beta<\alpha<1$ then $v(p)$ is monotone decreasing on $[0,(\alpha+\beta-1)/2]$ and
monotone increasing on $[(\alpha+\beta-1)/2,\alpha+\beta-1]$.
\item If $0<\alpha=\beta<1$ then $v(p)$ is constantly equal to $1$.
\end{itemize}
\end{Theorem}
\textbf{Sketch of the Proof}\quad
Basically we just have to consider the fluctuations of $v(p)$ and $W(p)$ using elementary calculus. \qed
The following corollary follows immediately.
\begin{Corollary}
We have
\begin{itemize}
\item $v(p) \gtreqqless v(q)\Longleftrightarrow W(p) \gtreqqless W(q)$ $($ if $\alpha>\beta)$,
\item $v(p) \gtreqqless v(q)\Longleftrightarrow W(p) \lesseqqgtr W(q)$ $($ if $\alpha<\beta)$.
\end{itemize}
between the velocity and the amplitude of one-soliton solution of \eqref{YBdKdV}.
\end{Corollary}
Therefore if $\alpha<\beta$ the smaller soliton moves faster than the larger one in $N$-soliton solutions.
For example, if we consider the two-soliton solution with parameters $\alpha=5/6,\ \beta=14/15,\ p_1=2/15,\ \gamma_1=-1/6,\ p_2=1/30,\ \gamma_2=-1/30$, then we obtain
\begin{eqnarray*}
v(p_1)=\frac{\log\frac{8}{3}}{\log\frac{7}{2}}=0.783 &>& v(p_2)=\frac{\log\frac{9}{2}}{\log 8}=0.723,\\
W(p_1)=0.363 &<& W(p_2)=0.722.
\end{eqnarray*}
Figure \ref{figure1} shows the two-soliton solution described above.
Lastly we note on the case $\alpha=\beta$.
In this case, clearly every solution has a speed $1$ everywhere, since \eqref{YBdKdV} becomes
just $x_n^{t+1}=y_n^t$ and $y_{n+1}^t=x_n^t$.  
In fig. \ref{figure2} we show a two-soliton solution where the parameters are taken as $\alpha=\beta=5/6,\ p_1=1/15,\ \gamma_1=-20,\ p_2=1/30,\ \gamma_2=-1/60$.

Note that as we have imposed $0<\alpha,\ \beta<1$ in this section, we cannot take a limit $a\to +\infty$ and obtain a solution of the normal discrete KdV eq. \eqref{dKdV2}. Hence the solutions obtained here are unique to the eq. \eqref{YBdKdV}.
\section{Ultradiscrete Limit of the Generalized Discrete KdV Equation}
Next we take the ultradiscrete limit of the generalized discrete KdV eq. \eqref{YBdKdV} and the $N$-soliton solutions. We show that the ultradiscrete limit of \eqref{YBdKdV} gives a box ball system with a carrier (BBSC).
\begin{Proposition}
The ultradiscretization of \eqref{YBdKdV} is equivalent to BBSC \eqref{BBSC}.
\end{Proposition}
\textbf{Proof}
We define $X_n^t,\ Y_n^t,\ A,\ B$ by
\begin{align*}
x_n^t&=\exp\left(-\frac{X_n^t}{\epsilon}\right),\ y_n^t=\exp\left(-\frac{Y_n^t}{\epsilon}\right),\\
\alpha&=\exp\left(-\frac{A}{\epsilon}\right),\ \beta=\exp\left(-\frac{B}{\epsilon}\right),
\end{align*}
where $\epsilon>0$ is an arbitrary parameter,
and then take the ultradiscrete limit of \eqref{YBdKdV}, which means that we take $\lim_{\epsilon\to +0}\epsilon \log (\cdot )$ of both sides of the eq. \eqref{YBdKdV}.
From the first eq. of \eqref{YBdKdV} we obtain
\begin{align*}
X_n^{t+1}&=\min\left[0,\ B+X_n^t+Y_n^t\right]+Y_n^t-\min\left[0,\ A+X_n^t+Y_n^t\right]\\
&=Y_n^t+\min\left[0,\ B+X_n^t+Y_n^t\right]+\max\left[0,\ -A-X_n^t-Y_n^t\right]\\
&=\min\left[-X_n^t,\ B+Y_n^t\right]+X_n^t+\max\left[Y_n^t,\ -A-X_n^t\right]\\
&=\min[-X_n^t,\ B+Y_n^t]+\max\left[X_n^t+Y_n^t+A,\ 0\right]-A.
\end{align*}
Note that since $0<\alpha<1$, we have $A>0$ and $\lim_{\epsilon\to+0}\epsilon\log\left(1-\exp\left(-\frac{A}{\epsilon}\right)\right)=0$ follows.
The same fact applies to $\beta$.
The relation $x_{n}^{t+1} y_{n+1}^t=x_n^t y_n^t$ shows that $Y_{n+1}^t=X_n^t+Y_n^t-X_n^{t+1}$.
If we displace $X_n^t$ and $Y_n^t$ as
\begin{equation}
U_n^t=X_n^t+A,\ \ \ V_n^t=Y_n^t+B, \label{udtransf}
\end{equation}
we obtain the following equation 
\begin{align*}
\left\{
\begin{array}{rl}
U_n^{t+1}=&\min\left[A-U_n^t,\ V_n^t\right]+\max\left[0,\ U_n^t+V_n^t-B\right],\vspace{2mm}\\
V_{n+1}^t=&U_n^t+V_n^t-U_n^{t+1},
\end{array}
\right.
\end{align*}
which is exactly the BBSC \eqref{BBSC} itself with $C_B=A$ and $C_C=B$.\qed
Note that transformations \eqref{restr} between the discrete systems correspond to \eqref{udtransf} in the ultradiscrete case and that the ultradiscrete limit of the Yang-Baxter map \eqref{YB2}
is the original BBSC \eqref{BBSC}.
\begin{Theorem}
We have the relation
\[
\beta  \gtreqqless \alpha \Longleftrightarrow C_B  \gtreqqless C_C
\]
between the parameters of the generalized discrete KdV eq. \eqref{YBdKdV} and those of the BBSC \eqref{BBSC}
when $0<\alpha,\ \beta<1$.
\end{Theorem}
\textbf{Proof} $\alpha<\beta$ is equivalent to $C_C=B<A=C_B$. The rest of the theorem is shown in the same manner.
\qed
Thus the inequality relation between the parameters  $\alpha$ and $\beta$ of generalized discrete KdV equation corresponds exactly to that of the parameters $C_B$ and $C_C$ of BBSC.  The following example illustrates
the soliton solutions of the BBSC when $C_C<C_B$.
\begin{figure}
\centering
\includegraphics[width=8cm,bb=150 -150 640 480]{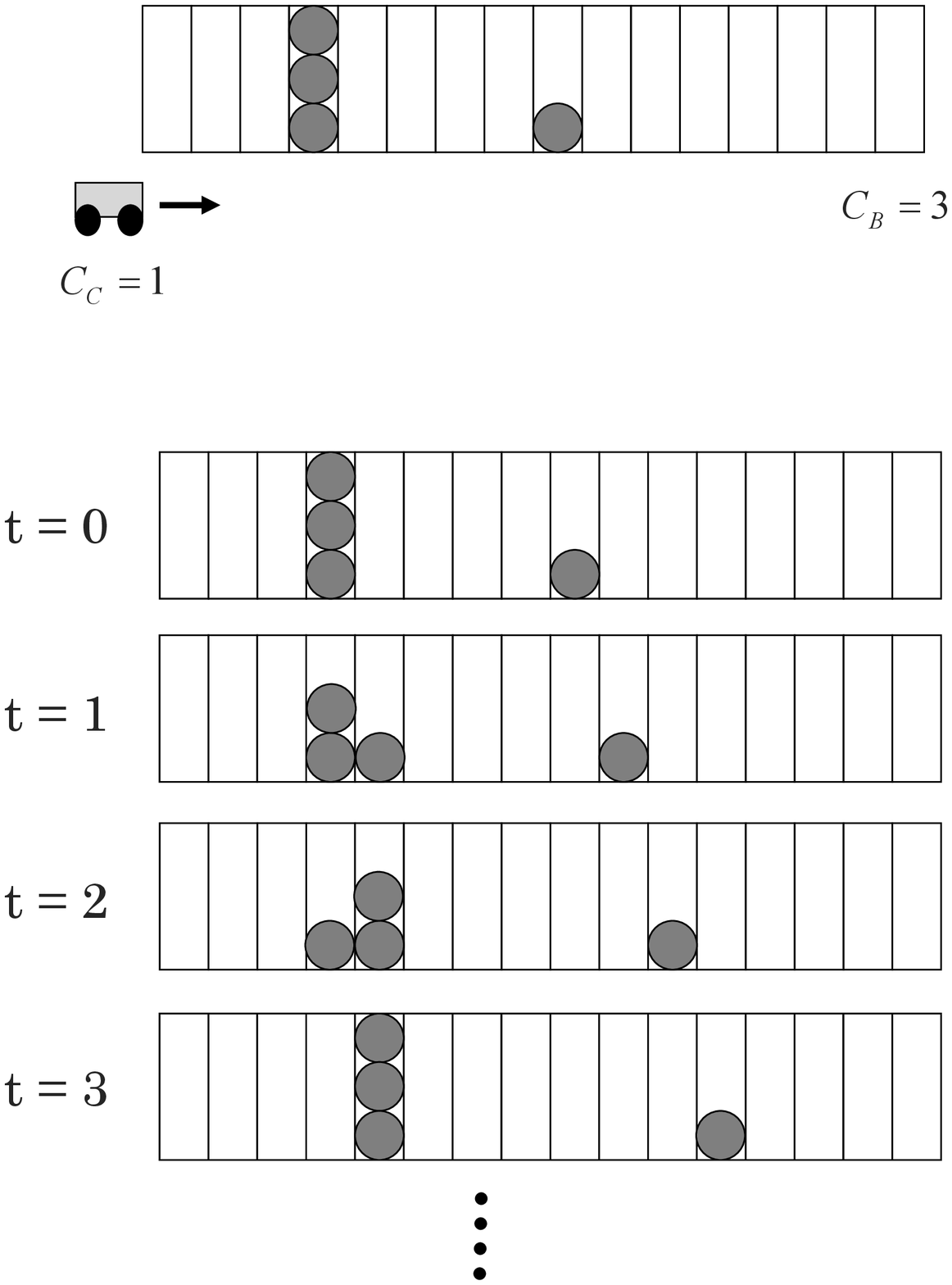}
\caption{The time evolution of the BBSC for $C_B=3$ and $C_C=1$.}
\label{figure3}
\end{figure}
The fig. \ref{figure3} shows the time evolution of $U_n^t$ if $C_B=3$ and $C_C=1$. The speed of the soliton on the left (larger) is $\dfrac{1}{3}$ and that of the soliton on the right (smaller) is $1$.
The reason why the smaller soliton moves faster is clear;
the capacity of the carrier is just one and it can carry a soliton with amplitude 1 at a time.
For a soliton with amplitude 3, the carrier cannot carry it all at once and it takes three time steps to move them to the next box. 

We conclude that these phenomena in soliton solutions of the BBSC coincide with those in the generalized discrete KdV equation described in fig. \ref{figure1}.
\section{Concluding Remarks}
We have obtained the velocity and the amplitude of soliton solutions to the generalised discrete KdV equation.
We have found the cases in which the smaller solitons take over the larger ones.
Through the ultradiscrete limit, these soliton solutions turn to the solutions of the BBSC. 
In BBSC, larger cluster of balls can move slower than the smaller ones, which corresponds to the phenomena we have obtained for the discrete equation in this article.
Detailed analysis of the solutions of other discrete integrable equations and of its ultradiscrete limits are the problems we would like to address in the future. 

\section*{Acknowledgment}
The authors wish to thank Professors Yasuhiro Ohta, Junkichi Satsuma and Ralph Willox for useful comments.


\end{document}